\documentclass[showpacs,amsmath,amssymb,twocolumn,nofootinbib]{revtex4}
\usepackage{epsfig}
\input{epsf}
\usepackage{amssymb}
\mathchardef\SGamma="7100
\begin{document}
\title{Thermodynamics via Creation from Nothing: Limiting
the Cosmological Constant Landscape\footnote{\normalsize To the
memory of Bryce DeWitt.}}

\author{A.O.Barvinsky}
\affiliation{Theory Department, Lebedev
Physics Institute, Leninsky Prospect 53, 119991 Moscow, Russia}
\author{A.Yu.Kamenshchik}
\affiliation{Dipartimento di Fisica and INFN, Via
Irnerio 46, 40126 Bologna, Italy\\
L.D. Landau Institute for Theoretical Physics, Kosygin str. 2,
119334 Moscow, Russia}
\begin{abstract}
The creation of a quantum Universe is described by a {\em density
matrix} which yields an ensemble of universes with the cosmological
constant limited to a bounded range $\Lambda_{\rm min}\leq \Lambda
\leq \Lambda_{\rm max}$. The domain $\Lambda<\Lambda_{\rm min}$ is
ruled out by a cosmological bootstrap requirement (the
self-consistent back reaction of hot matter). The upper cutoff
results from the quantum effects of vacuum energy and the conformal
anomaly mediated by a special ghost-avoidance renormalization. The
cutoff $\Lambda_{\rm max}$ establishes a new quantum scale -- the
accumulation point of an infinite sequence of garland-type
instantons. The dependence of the cosmological constant range on
particle phenomenology suggests a possible dynamical selection
mechanism for the landscape of string vacua.
\end{abstract}
\maketitle

Quantum cosmology \cite{Bryce,VilNB} and Euclidean quantum gravity
\cite{HH} might effectively restrict the landscape of string vacua.
This landscape is too big \cite{BoussoPolchinsky} to predict either
the observed particle phenomenology or large-scale structure
formation within string theory itself. Other methods have to be
invoked, at least some of them based on the cosmological
wavefunction \cite{OoguriVafaVerlinde,Tye,Brustein}. This is
dominated by the exponentiated Euclidean action, $\exp(-S_{\rm E})$,
calculated on the gravitational instanton which is a saddle point of
an underlying path integral over Euclidean 4-geometries. This
instanton gives rise to Lorentzian signature spacetime by analytic
continuation across minimal hypersurfaces. The continuation can be
interpreted either as quantum tunneling or as the creation of the
Universe from ``nothing". Thus, the most probable vacua of the
landscape become weighted by the minima of $S_{\rm E}$. This might
serve as a method of selecting a vacuum from the enormously big
string landscape.

An immediate difficulty with this program arises from the infrared
catastrophe of small cosmological constant $\Lambda$. The
Hartle-Hawking wave function \cite{HH}, which describes nucleation
of the de Sitter Universe from the Euclidean 4-dimensional
hemisphere, has the form
    \begin{eqnarray}
    \Psi_{\rm HH}\sim \exp(-S_{\rm E})
    =\exp(3\pi /2G\Lambda).                         \label{1}
    \end{eqnarray}
This diverges for $\Lambda\to 0$ because of unboundedness of the
Euclidean gravitational action. Despite some early attempts to
interpret it as the origin of a zero value of $\Lambda$
\cite{bigfix}, this result remains both controversial and
anti-intuitive because it disfavors inflation and prefers creation
of infinitely large universes. Apart from the tunneling proposals of
\cite{tunnel} which employ an opposite sign in the exponential of
(\ref{1}) and thus open the possibility for opposite conclusions
\cite{scale}, no convincing resolution of this problem has thus far
been suggested.

In this Letter we show that Euclidean path integration framework
naturally avoids this infrared catastrophe. We attain this result
by: i) extending the notion of Hartle-Hawking {\em pure} state to a
density matrix which describes a {\em mixed} quantum state of the
Universe and ii) incorporating the nonperturbative back reaction of
hot quantum matter on the instanton background \cite{slih}. These
extensions seem natural because whether the initial state of the
Universe is pure or mixed is a dynamical question rather than a
postulate. We address this question below by embedding both types of
states into a unified framework of a density matrix.
\begin{figure}[h]
\centerline{\epsfxsize 4.4cm \epsfbox{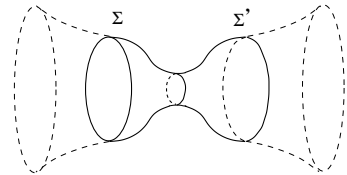}} \caption{\small
Density matrix instanton. Dashed lines depict the Lorentzian
Universe nucleating at minimal surfaces $\Sigma$ and $\Sigma'$.
\label{Fig.1}}
\end{figure}

\begin{figure}[h]
\centerline{\epsfxsize 4.3cm \epsfbox{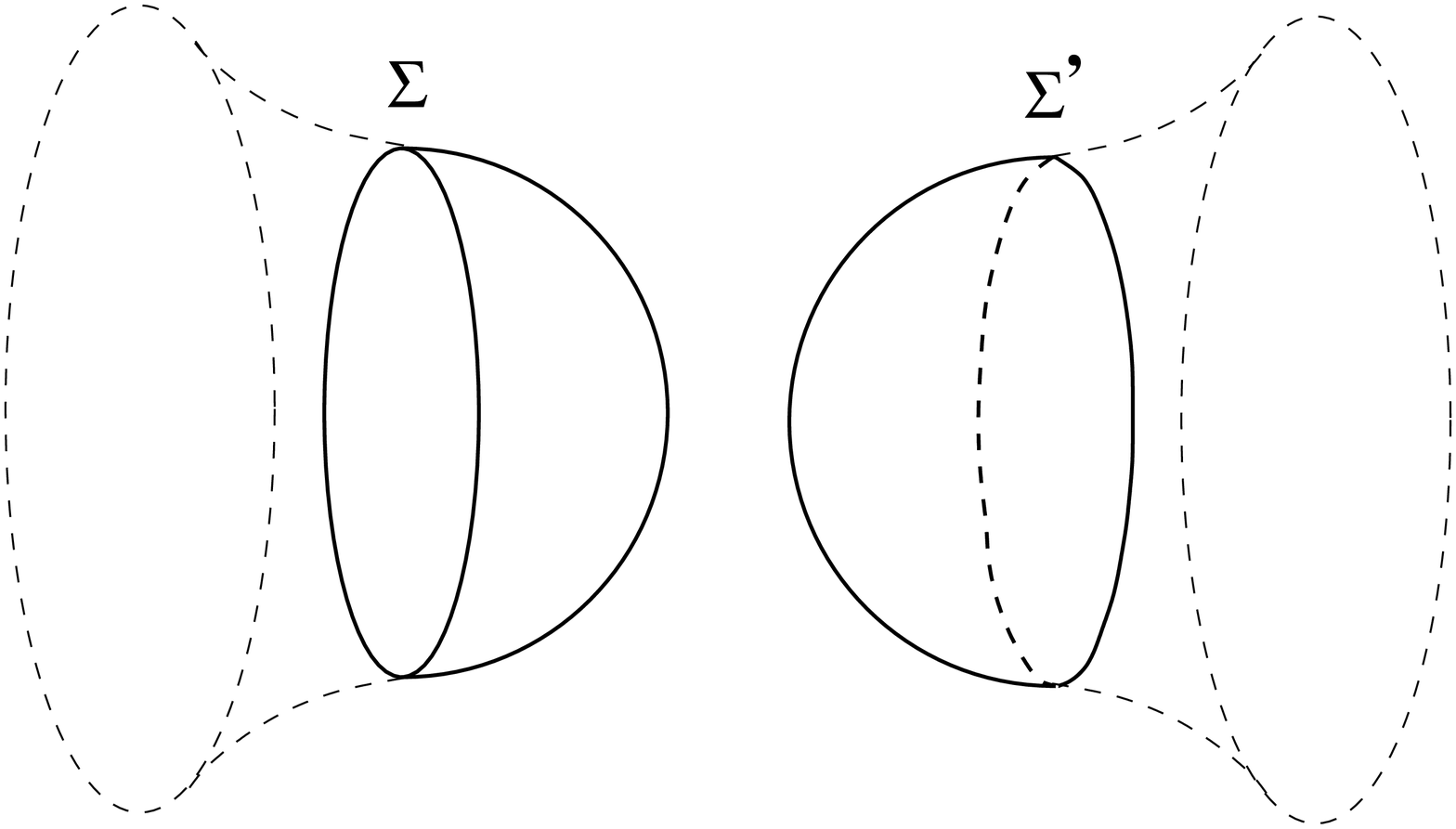}} \caption{\small
Density matrix of the pure Hartle-Hawking state represented by the
union of two vacuum instantons. \label{Fig.2}}
\end{figure}

A density matrix $\rho[\,\varphi,\varphi']$ is represented in
Euclidean quantum gravity \cite{Page} by an instanton having two
disjoint boundaries $\Sigma$ and $\Sigma'$ associated with its two
entries $\varphi$ and $\varphi'$ (collecting both gravity and matter
observables). The instanton interpolates between these, thus
establishing mixing correlations, see Fig.\ref{Fig.1}. In contrast,
for the density matrix of the pure Hartle-Hawking state the bridge
between $\Sigma$ and $\Sigma'$ is broken, so that the instanton is a
union of two disjoint hemispheres which smoothly close up at their
poles (Fig.\ref{Fig.2}) --- a picture illustrating the factorization
of $\hat\rho=|\Psi_{\rm HH}\rangle\langle\Psi_{\rm HH}|$.

The main effect that we advocate here is that thermal fluctuations
and quantum conformal anomaly destroy the Hartle-Hawking instanton
and replace it with one filled by radiation. This is already
manifest in classical theory, specifically in the Euclidean
Friedmann equation for a scale factor $a(\tau)$,
$\dot{a}^2/a^2=1/a^2 - H^2 -C/a^4$ (we use spatially closed FRW
metric $ds^2 = N^2(\tau)\,d\tau^2 +a^2(\tau)\,d^2\Omega^{(3)}$ in
the gauge $N=1$ and express $\Lambda=3H^2$ in terms of the Hubble
constant). The radiation density $C/a^4$ prevents the
half-instantons from closing and allows $a$ to vary between two
turning points \cite{Halliwell,Rubakov}
    $a_\pm=(1/\sqrt{2}H)(1\pm\sqrt{1-4CH^2})^{1/2}$.
This forces a tubular structure on the instanton which spans (at
least) one period of oscillation between $a_\pm$, provided the
constant $C$ characterizing the amount of radiation satisfies the
bound $4H^2C \leq 1$.

The existence of radiation naturally follows from the partition
function of this state associated with the toroidal instanton
obtained by identifying $\Sigma'$ and $\Sigma$. The radiation back
reaction supports the instanton geometry in which it exists.
Remarkably, when the vacuum energy and conformal anomaly are taken
into account this bootstrap yields a set of instantons -- a
landscape -- only in the bounded range of $\Lambda$,
    \begin{eqnarray}
    \Lambda_{\rm min}<\Lambda<\Lambda_{\rm max}.            \label{2}
    \end{eqnarray}
All values $\Lambda<\Lambda_{\rm min}$ are completely eliminated
either because of the absence of instanton solutions or because of
their {\em infinitely large positive} action. A similar situation
holds for $\Lambda>\Lambda_{\rm max}$ -- no instantons exist there,
and the Lorentzian configurations in this overbarrier domain (if
any) are exponentially suppressed relative to those of (\ref{2}).

To quantify the above picture consider the density matrix given by
the Euclidean path integral \cite{Page}
    \begin{eqnarray}
    \rho[\,\varphi,\varphi'\,]=
    \mbox{\large$e$}^{\,\textstyle \varGamma}
    \int
    D[\,g,\phi\,]\,
    \exp\big(-S_{\rm E}[\,g,\phi\,]\big),             \label{rho}
    \end{eqnarray}
where $S_{\rm E}[\,g,\phi\,]$ is the classical action, and the
integration runs over gravitational $g$ and matter $\phi$ fields
interpolating between $\varphi$ and $\varphi'$ at $\Sigma$ and
$\Sigma'$. The statistical sum $\exp(-\varGamma)$ is given by a
similar path integral over periodic fields on the torus with
identified boundaries $\Sigma$ and $\Sigma'$.

The back reaction follows from decomposing $[g,\phi]$ into a
minisuperspace $g_0(\tau)=\big(a(\tau),N(\tau)\big)$, and the
"matter" sector which includes also inhomogeneous metric
perturbations on minisuperspace background
$\varPhi(x)=(\phi(x),\psi(x),A_\mu(x), h_{\mu\nu}(x),...)$. With a
relevant decomposition of the measure $D[\,g,\phi\,]=Dg_0(\tau)
\times D\varPhi(x)$, the integral for $\varGamma$ expresses in terms
of the effective action $\varGamma[g_0(\tau)]$ of quantized matter
on the background $g_0(\tau)$,
    $\exp(-\varGamma[g_0])=
    \int D\varPhi(x)\,
    \exp(\!-S_{\rm E}[g_0,\varPhi(x)])$,
as
    \begin{eqnarray}
    \mbox{\large$e$}^{\textstyle
    -\SGamma}=
    \int Dg_0(\tau)\,
    \exp\Big(\!-\varGamma[g_0(\tau)]\Big).          \label{intg0}
    \end{eqnarray}
Our approximation will be to consider $\varGamma[g_0(\tau)]$ in the
one-loop order, $\varGamma[g_0]=S_{\rm E}[g_0]+\varGamma_{\rm
1-loop}[g_0]$, and handle (\ref{intg0}) at the tree level, which is
equivalent to solving the {\em effective equations} for
$\varGamma[g_0]$.

Remarkably, $\varGamma[\,g_0\,]$ is {\em exactly} calculable for
confor\-mally-invariant fields by a conformal transformation
\cite{FHH} relating generic FRW metric $ds^2 = a^2 d\bar s^2$ to
that of a static universe of a unit size, $d\bar s^2 = d\eta^2 +
d^2\Omega^{(3)}$ (these metrics are denoted below as $g$ and $\bar
g$, while $\eta$ is the conformal time). The total action reads
    \begin{eqnarray}
    &&\varGamma[\,a(\tau),N(\tau)\,]=
    2 \int_{\tau_-}^{\tau_+} d\tau\Big(\!-\frac{a\dot{a}^2}N
    -Na+N H^2 a^3\Big)\nonumber\\
    &&\qquad\qquad
    +2B \int_{\tau_-}^{\tau_+}
    d\tau \Big(\frac{\dot{a}^2}{Na}
    -\frac16\,\frac{\dot{a}^4}{N^3 a}\Big)\nonumber\\
    &&\qquad\qquad
    + B \int_{\tau_-}^{\tau_+}
    d\tau\,N/a+F\Big(2\int_{\tau_-}^{\tau_+}
    d\tau\,N/a\Big)\,.                      \label{Gamma}
    \end{eqnarray}
We work in units of $m_P=\sqrt{3\pi/4G}$, and the integration runs
between two turning points at $\tau_\pm$. The first line is the
classical part, the second line is the conformal contribution and
the last line is the one-loop action on the static instanton of the
metric $\bar g$.

The conformal contribution $\varGamma_{\rm 1-loop}[g]-\varGamma_{\rm
1-loop}[\bar g]$ is determined by the coefficients of $\Box R$, the
Gauss-Bonnet invariant $E = R_{\mu\nu\alpha\gamma}^2 -4R_{\mu\nu}^2
+ R^2$ and Weyl tensor term in the conformal anomaly
    $g_{\mu\nu}\delta
    \SGamma_{\rm 1-loop}/\delta g_{\mu\nu} =
    g^{1/2}
    (\alpha \Box R +
    \beta E + \gamma C_{\mu\nu\alpha\beta}^2)/4(4\pi)^2 $.
Specifically this contribution can be obtained by the technique of
\cite{TseytlinconfBMZ}; it contains higher-derivative terms $\sim
\ddot a^2$ which produce ghost instabilities in solutions of
effective equations. However, such terms are proportional to the
coefficient $\alpha$ which can be put to zero by adding the
following finite {\em local} counterterm
    \begin{eqnarray}
    \varGamma_{R}[g]
    =\varGamma_{\rm 1-loop}[g]
    +\frac1{2(4\pi)^2}
    \frac\alpha{12}\int d^4x\,
    g^{1/2}R^2(g).                   \label{renormalization}
    \end{eqnarray}
This ghost-avoidance renormalization is justified by the requirement
of consistency of the theory at the quantum level. The contribution
$\varGamma_{R}[g]-\varGamma_{R}[\bar g]$ to the {\em renormalized}
action then gives the second line of (\ref{Gamma}) with
$B=3\beta/4$.

The static instanton with a period $\eta_0$ playing the role of
inverse temperature contributes $\varGamma_{\rm 1-loop}[\bar g]
=E_0\,\eta_0+F(\eta_0)$, where the vacuum energy $E_0$ and free
energy $F(\eta_0)$ are the typical boson and fermion sums over field
oscillators with energies $\omega$ on a unit 3-sphere
    $E_0=\pm\sum_{\omega}
    \omega/2\,,\,\,\,F(\eta_0)=\pm\sum_{\omega}
    \ln\big(1\mp e^{-\omega\eta_0}\big)$.
The renormalization (\ref{renormalization}) which should be applied
also to $\varGamma_{\rm 1-loop}[\bar g]$ modifies $E_0$, so that
$\varGamma_{R}[\bar g] =C_0\,\eta_0+F(\eta_0)$, $C_0\equiv
E_0+3\alpha/16$. This gives the third line of Eq.(\ref{Gamma}) with
$C_0=B/2$. This universal relation between $C_0$ and $B=3\beta/4$
follows from the known anomaly coefficients \cite{confanomaly} and
the UV-renormalized Casimir energy in a static universe \cite{E_0}
for scalar, Weyl spinor and vector fields respectively having:
    \begin{eqnarray}
    \alpha=\frac1{90}\!\times\!\left\{\begin{array}{c} \!-1 \\
    \!-3\\
    \!18\end{array}\right.\!\!,
    \,\,
    \beta=\frac1{360}\!\times\!
    \left\{\begin{array}{cl} \!\!2\\
    \!\!11\\
    \!\!124\end{array}\right.\!\!\!\!,
    \,\,
    E_0=\frac1{960}\!\times\!\left\{\begin{array}{c} \!\!4 \\
    \!17\\
    \!88\end{array}\right.\!\!\!.       \label{5000}
    \end{eqnarray}
It is important that for conformally invariant fields the nonlocal
action (\ref{Gamma}) is exact, and contains no other terms of higher
order in the curvature.

The effective equation $\delta\varGamma/\delta N(\tau)=0$ has the
form of the classical equation modified by the quantum $B$-term
    \begin{eqnarray}
    &&\frac{\dot{a}^2}{a^2}
    +B \left(\frac12\,\frac{\dot{a}^4}{a^4}
    -\frac{\dot{a}^2}{a^4}\right) =
    \frac{1}{a^2} - H^2 -\frac{C}{ a^4},     \label{efeq}\\
    &&C = B/2 +F'(\eta_0),\,\,
    \eta_0 = 2\int_{\tau_-}^{\tau_+}
    d\tau/a(\tau).                       \label{bootstrap}
    \end{eqnarray}
Remarkably, the contribution of the nonlocal $F(\eta_0)$ in
(\ref{Gamma}) reduces to the radiation {\em constant} $C$ as a {\em
nonlocal functional} of $a(\tau)$, determined by the {\em bootstrap}
equation (\ref{bootstrap}). Here $F'(\eta_0)\equiv
dF(\eta_0)/d\eta_0>0$ is the energy of a hot gas of particles, which
adds to their vacuum energy $B/2$.

\begin{figure}[h]
\centerline{\epsfxsize 8.5cm \epsfbox{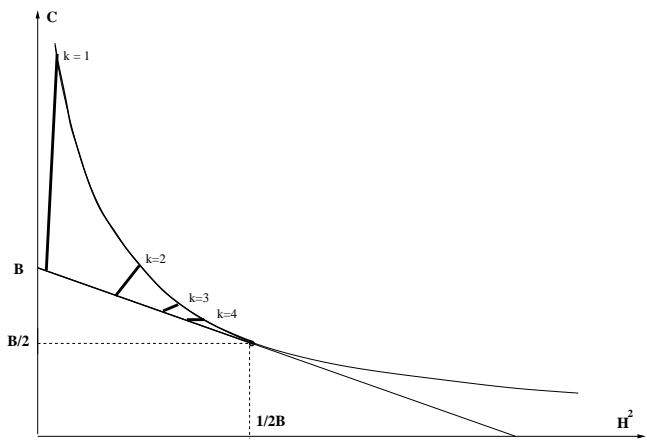}} \caption{\small
Instanton domain in the $(H^2,C)$-plane. Garland families are shown
for $k=1,2,3,4$. Their sequence accumulates at the critical point
$(1/2B,B/2)$.
 \label{Fig.4}}
\end{figure}
Periodic instanton solutions of Eqs.(\ref{efeq})-(\ref{bootstrap})
exist only inside the curvilinear wedge of $(H^2,C)$-plane between
bold segments of the upper hyperbolic boundary and the lower
straight line boundary of Fig.\ref{Fig.4},
    \begin{equation}
    4CH^2\leq 1,\,\,\,C \geq B-B^2 H^2,
    \,\,\,B H^2\leq 1/2.             \label{restriction1}
    \end{equation}
Below this domain the solutions are either complex and aperiodic or
suppressed by {\em infinite positive} Euclidean action. Above this
domain only Lorentzian (overbarrier) configurations exist, but they
are again exponentially damped relative to instantons in
(\ref{restriction1}).

These properties are based on the fact that the turning points of
(\ref{efeq}) exactly coincide with classical $a_\pm$, but $a_-$
exists only when $a_-^2 \geq B$, which gives rise to
(\ref{restriction1}). Otherwise, $a(\tau)$ at the contraction phase
becomes complex or runs to zero which violates instanton
periodicity. In the latter case a smooth Hartle-Hawking instanton
with $a_-=0$ forms and yields  $\eta_0\to\infty$ in view of
(\ref{bootstrap}), so that $F(\eta_0)\sim F'(\eta_0)\to 0$.
Therefore, its on-shell action
    \begin{equation}
    \varGamma_0= F(\eta_0)\!-\!\eta_0 F'(\eta_0)
    +4\!\int_{a_-}^{a_+}\!
    \frac{da \dot{a}}{a}\Big(B-a^2
    -\frac{B\dot{a}^2}{3}\Big)              \label{action-instanton}
    \end{equation}
due to $B>0$ diverges to $+\infty$ at $a_-=0$ and completely rules
out pure-state instantons \cite{slih}.

Moreover, inside the range (\ref{restriction1}) our bootstrap
eliminates the infrared catastrophe of $\Lambda\to 0$. Indeed
$\eta_0\to\infty$ as $H^2\to 0$, so that due to (\ref{bootstrap})
$C\to B/2$, but this is impossible because in view of
(\ref{restriction1}) $C\geq B$ at $H^2=0$. Thus, instanton family
never hits the $C$-axes of $H^2=0$ and can only interpolate between
the points on the boundaries of the domain (\ref{restriction1}). For
a conformal scalar field the numerical analysis gives such a family
\cite{slih} starting at $H^2\approx2.00,\,\,C\approx0.004,\,\,\,
\varGamma_0\approx-0.16$, and terminating at
$H^2\approx13.0,\,\,C\approx0.02 ,\,\,\,\varGamma_0\approx-0.09$.
The upper point describes the static universe filled by a hot
radiation with the temperature $T=H/\pi\sqrt{1-2BH^2}$, whereas the
lower point establishes the lower bound of the $\Lambda$-range.

\begin{figure}[h]
\centerline{\epsfxsize 5cm \epsfbox{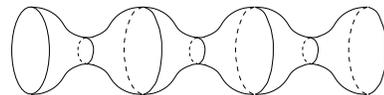}} \caption{\small The
garland segment consisting of three folds of a simple instanton
glued at surfaces of a maximal scale factor.
 \label{Fig.5}}
\end{figure}

The upper bound of the landscape follows from the existence of {\em
garlands} that can be obtained by glueing together into a torus $k$
copies of a simple instanton \cite{Halliwell,mult-inst}; see
Fig.\ref{Fig.5}. Their formalism is the same as above except that
the conformal time in (\ref{bootstrap}) and the integral term of
(\ref{action-instanton}) should be multiplied by $k$.
As in the case of $k=1$, garland families interpolate between the
lower and upper boundaries of (\ref{restriction1}). They exist for
all $k$, $1\leq k\leq\infty$, and their infinite sequence
accumulates at the critical point $C=B/2$, $H^2=1/2B$, where these
boundaries merge. Within the $1/k^2$-accuracy the upper and lower
points of each family coincide and read
    \begin{eqnarray}
    H^2_{(k)}\simeq \frac1{2B}
    \left(1 - \frac{\ln^2k^2}{2k^2\pi^2}
    \right),\,\,\,\varGamma_0^{(k)}\simeq
    -B\,\frac{\ln^3 k^2}{4k^2\pi^2}.         \label{action-k}
    \end{eqnarray}
With a growing $k$, garlands become more and more static and cold
with $T_{(k)}\simeq 1/(\sqrt B \ln k^2)\to 0$. Contrary to
\cite{mult-inst} the garland action is not additive in $k$, so that
as $k\to\infty$ garlands do not dominate the ensemble. Their
sequence converges to the instanton with $H^2_{\rm max}=1/2B$, which
gives the upper bound of the range (\ref{2}).

Thus, our Universe is created in a hot mixed state, but its
evolution does not contradict the large-scale structure formation.
After nucleation from the instanton the Universe expands; its
radiation dilutes, so that $\Lambda$ starts dominating and generates
inflation under an assumption that everywhere above $\Lambda$ is a
composite field (like an inflaton) decaying at the exit by a
standard slow-roll scenario.

The ensemble of universes belongs to a bounded range (\ref{2}) of
$\Lambda=3H^2$. Its infrared cutoff is provided by the radiation
back reaction and survives even in the classical limit as $B\to 0$.
In contrast, the high-energy cutoff at
    \begin{eqnarray}
    \Lambda_{\rm max}=3m_P^2/2B,\,\,\,
    m_P^2\equiv 3\pi/4G,              \label{top}
    \end{eqnarray}
is the quantum effect of vacuum energy and the conformal anomaly;
this generates a new scale in gravity theory.

We have considered only conformal fields which make our model
exactly solvable and provide critically important positivity of the
constant $B=3\beta/4$, cf. (\ref{5000}). Moreover, conformal
invariance together with the ghost-avoidance renormalization renders
a particular value of the vacuum energy $B/2$ in (\ref{bootstrap})
which yields the upper boundary of (\ref{2}) exactly at the critical
point $(1/2B,B/2)$ of Fig.\ref{Fig.4}. Even if non-conformal fields
qualitatively preserve the whole picture, they are likely to break
this relation. Then if $C_0<B/2$ all garlands survive, though they
saturate at $\Lambda_{\rm max}$ with a finite temperature. If
$B>C_0>B/2$, their sequence is truncated at some $k$. Finally, if
$C_0>B$ the infrared catastrophe occurs again -- the $k=1$ instanton
family hits the $C$-axes at $C_0$.

Conformal invariance can be justified as a good approximation when
conformal particles outnumber non-conformal ones. Moreover, their
large number $N$ justifies a semiclassical expansion by scaling down
the range (\ref{2}). Indeed, for a single scalar field the latter is
determined by Planckian values, $\Lambda_{\rm min}\approx 8.99\,
m_P^2$, $\Lambda_{\rm max}=360\, m_P^2$ which, however, decrease as
$1/N$ in view of the simple scaling $C\to NC$, $B\to NB$,
$F(\eta_0)\to NF(\eta_0)$ and $H^2\to H^2/N$. Semiclassical
expansion can also be justified for large $B=3\beta/4$ growing with
spin, cf.(\ref{5000}), because the domain (\ref{2}) with (\ref{top})
shrinks to a narrow subplanckian range when ascending the particle
hierarchy.

Though motivated by the string landscape, all the above results hold
outside of the string theory context and, as a feedback, suggest a
long-sought selection mechanism for the plethora of string vacua.
Modulo the details of a relevant $4D$-compactification, this might
work as follows. For $B$ growing with $N$ and spin, the upper scale
(\ref{top}) decreases towards the increasing phenomenology scale,
and approaches the latter at the string scale $m_{\rm s}^2$ where a
positive $\Lambda$ might be generated by the mechanism of
\cite{KKLT}. Our conjecture is that at this scale our bootstrap
becomes perturbatively consistent, provided $m_P^2/B\simeq m_{\rm
s}^2\ll m_P^2$, and selects from the string landscape a small subset
compatible with observations.

Our results hold within the Euclidean path integral (\ref{rho})
which automatically excludes Lorentzian configurations possibly
existing above the upper boundary of (\ref{restriction1}),
$4CH^2>1$. However, one can imagine an extended formulation of
quantum gravity generalizing (\ref{rho}) to a wider path integration
domain. Our conclusions nevertheless remain true. Indeed the
effective action scales as $\varGamma_0\!\sim\!-\sqrt B$, $B\gg 1$,
and because it is {\em negative} our landscape at the scale $m_{\rm
s}$ is weighted by $\exp(\#\sqrt B)\!\simeq\!\exp(\#\,m_P/m_{\rm
s})\gg 1$. Therefore it strongly dominates over Lorentzian
configurations, the amplitudes of the latter being $O(1)$ in view of
their pure phase nature. Thus, our results look robust against
possible generalizations of Euclidean quantum gravity.\\

A.B. thanks H.Tye, C.Deffayet, J.Khoury, R.Woodard, T.Osborn and
especially Jim Hartle for thought provoking discussions and helpful
suggestions. A.B. was supported by the RFBR grant 05-02-17661 and
the grant LSS 4401.2006.2. A.K. was partially supported by the RFBR
grant 05-02-17450 and the grant LSS 1757.2006.2.

\end{document}